\documentclass[aps,pra,reprint,superscriptaddress,floatfix,nofootinbib]{revtex4-2}

\usepackage[english]{babel}
\usepackage[utf8]{inputenc}
\usepackage[T1]{fontenc}

\usepackage{amsfonts,amssymb,amsmath,mathtools}
\usepackage{dsfont}
\usepackage{braket}
\usepackage{accents}
\usepackage{bm}

\usepackage{graphicx}
\usepackage{makecell, multirow}

\usepackage[colorlinks=true,citecolor=blue,linkcolor=blue,urlcolor=blue]{hyperref}
\usepackage[capitalise]{cleveref}

\newcommand{\ie}{i.\,e.}
\newcommand{\eg}{e.\,g.}
\newcommand{\wrt}{w.\,r.\,t.\;}
\newcommand{\mlx}{ML-X\;}
\DeclareMathOperator{\Tr}{Tr}

\usepackage{xcolor}
\newcommand{\colorOne}[1]{\textcolor{black}{#1}}
\newcommand{\colorTwo}[1]{\textcolor{black}{#1}}
\newcommand{\colorThree}[1]{\textcolor{black}{#1}}

\begin{document}
	
\title{Interaction-controlled impurity transport in trapped mixtures of ultracold bosons}

\author{Judith Becker}
\email{judith.becker@physnet.uni-hamburg.de}
\affiliation{Zentrum für Optische Quantentechnologien, Universität
	Hamburg, Luruper Chaussee 149, 22761 Hamburg, Germany}

\author{Maxim Pyzh}
\email{mpyzh@physnet.uni-hamburg.de}
\affiliation{Zentrum für Optische Quantentechnologien, Universität
	Hamburg, Luruper Chaussee 149, 22761 Hamburg, Germany}

\author{Peter Schmelcher}
\email{pschmelc@physnet.uni-hamburg.de}
\affiliation{Zentrum für Optische Quantentechnologien, Universität
	Hamburg, Luruper Chaussee 149, 22761 Hamburg, Germany}
\affiliation{The Hamburg Centre for Ultrafast Imaging, Universität 
	Hamburg, Luruper Chaussee 149, 22761 Hamburg, Germany}

\begin{abstract}
	We explore the dynamical transport of an impurity
	between different embedding majority species
	which are spatially separated in a double well.
	The transfer and storage of the impurity is triggered
	by dynamically changing the interaction strengths
	between the impurity and the two majority species.
	We find a simple but efficient protocol consisting  of 
	linear ramps of majority-impurity interactions
	at designated times to pin or unpin the impurity.
	Our study of this highly imbalanced few-body triple mixture
	is conducted with
	the multi-layer multi-configuration time-dependent Hartree method for atomic mixtures 
	which accounts for all interaction-induced correlations.
	We analyze the dynamics in terms of single-particle densities, 
	entanglement growth and provide an effective potential description 
	involving mean-fields of the interacting components.
	The majority components remain self-trapped in their individual wells at all times,
	which is a crucial element for the effectiveness of our protocol.				
	During storage times each component 
	performs low-amplitude dipole oscillations in a single well.
	Unexpectedly, the inter-species correlations 
	possess a stabilizing impact on the transport 
	and storage properties of the impurity particle.
\end{abstract}

\maketitle


\section{Introduction}\label{sec:intro}


Tunneling of microscopic particles through a classically forbidden barrier
is an exceptionally important quantum-mechanical phenomenon.
It is a direct consequence of the particle-wave duality
and the uncertainty principle.
Tunneling has a wide range of real-world applications:
it imposes a fundamental limit for the size of transistors 
\cite{kawaura_observation_2000}
and lies at the heart of numerous technological devices such as
the scanning tunneling microscope 
\cite{binnig_scanning_1987,wiesendanger_scanning_1994}, 
the tunnel diode \cite{esaki_new_1958},
ultrasensitive magnetometers (SQUID) \cite{jaklevic_quantum_1964} and
superconducting qubits \cite{kjaergaard_superconducting_2020}.
The concept has been used to explain fundamental problems 
in physics, chemistry and biology with great success, 
including radioactive decay processes \cite{gurney_wave_1928,gamow_zur_1928},
nuclear fusion \cite{balantekin_quantum_1998},
astrochemical synthesis \cite{trixler_quantum_2013},
chemical reactions \cite{bell_tunnel_2013} and DNA mutations 
\cite{lowdin_proton_1963}.


Tunneling can be observed on a macroscopic scale 
between two phase-coherent spatially overlapping matter waves \cite{josephson_possible_1962},
and has been detected via a current between two superconductors
separated by a thin insulating layer (SJJ), 
even though no external voltage is applied (dc-Josephson effect).
An external voltage gives then rise to a rapidly oscillating current
(ac-Josephson effect).
The Josephson effect 
\cite{likharev_superconducting_1979,barone_physics_1982} 
was also reported for superfluid helium 
\cite{pereverzev_quantum_1997,backhaus_direct_1997,
	sukhatme_observation_2001,davis_superfluid_2002},
cavity polaritons \cite{abbarchi_macroscopic_2013}
and ultracold atomic gases 
\cite{albiez_direct_2005,shin_optical_2005,
	schumm_matter-wave_2005,bouchoule_modulational_2005,
	gati_bosonic_2007,levy_c_2007,leblanc_dynamics_2011,betz_two-point_2011}.
The latter platform is of particular relevance 
for a quantitative analysis of the tunneling effect,
as it provides an exquisite control over system parameters 
and versatile detection techniques \cite{pethick_boseeinstein_2008}.
Atomic Josephson junctions \cite{leggett_bose-einstein_2001}
have been suggested as a standard of chemical potential
\cite{kohler_chemical_2003},
to perform measurements of gravity
\cite{hall_condensate_2007,baumgartner_measuring_2010} 
and off-diagonal long-range order \cite{ginsberg_coherent_2007}
with a high spatial resolution.


A double well loaded with a many-body ensemble of ultracold bosons, 
known as the bosonic Josephson junction (BJJ) 
\cite{javanainen_oscillatory_1986,dalfovo_order_1996,jack_coherent_1996}, 
has drawn particular attention
due to its fundamental nature and conceptual simplicity.
Two wells separated by a barrier is a paradigmatic external potential 
to investigate tunneling dynamics 
\cite{albiez_direct_2005,shin_optical_2005,
	schumm_matter-wave_2005,bouchoule_modulational_2005,
	gati_bosonic_2007,levy_c_2007,
	leblanc_dynamics_2011,betz_two-point_2011},
interference of matter waves 
\cite{andrews_observation_1997,orzel_squeezed_2001,
	schumm_matter-wave_2005,julia-diaz_dynamic_2012,
	kaufman_two-particle_2014},
the Shapiro \cite{eckardt_analog_2005,grond_shapiro_2011} 
and ratchet effects \cite{chen_asymptotic_2020},
macroscopic superposition states 
\cite{mahmud_quantum_2005,huang_creation_2006,dounas-frazer_ultracold_2007,
	garcia-march_macroscopic_2012,garcia-march_mesoscopic_2015} 
and entanglement 
\cite{bar-gill_einstein-podolsky-rosen_2011,
	he_einstein-podolsky-rosen_2011,kaufman_entangling_2015}.
Furthermore, it serves as a prototype model for finite-size lattices
\cite{anker_nonlinear_2005,winkler_repulsively_2006,
	folling_direct_2007,trotzky_time-resolved_2008,esteve_squeezing_2008}.


BJJ can be understood in a two-mode approximation 
(lowest band Bose-Hubbard model)
\cite{milburn_quantum_1997,smerzi_quantum_1997,zapata_josephson_1998,
	raghavan_coherent_1999,raghavan_transitions_1999,
	marino_bose-condensate_1999, smerzi_macroscopic_2000,
	giovanazzi_josephson_2000,javanainen_splitting_1999,
	ananikian_gross-pitaevskii_2006,salgueiro_quantum_2007}.
Non-interacting particles, initially prepared in one well, 
will perform Rabi oscillations between the two wells 
with a well-defined frequency.
For an ensemble of particles, 
the spectral response is strongly affected 
by tunable inter-particle interactions and the initial population imbalance,
evincing dc-/ac-Josephson effects and plasma oscillations 
\cite{smerzi_quantum_1997,zapata_josephson_1998,
	raghavan_coherent_1999,giovanazzi_josephson_2000}.
Moreover, interactions give rise to novel dynamical regimes, 
not possible with SJJ,
such as $\pi$-phase modes 
\cite{raghavan_transitions_1999,marino_bose-condensate_1999} and, 
above a critical value of the interaction strength, 
the macroscopic quantum self-trapping (MQST) 
\cite{milburn_quantum_1997,smerzi_quantum_1997}, 
i.e., a suppression of tunneling even though the particles repel each other.
Interestingly, the two-mode model has a classical analogue, 
namely it can be mapped to a non-rigid pendulum 
\cite{smerzi_quantum_1997,zapata_josephson_1998,
	raghavan_coherent_1999,marino_bose-condensate_1999,mahmud_quantum_2005}:
the population and relative-phase difference 
between two condensate fractions translate
to the angular momentum and displacement, respectively.
In particular, MQST corresponds to the pendulum making full revolutions, 
implying a non-zero average population imbalance
and the relative phase increasing monotonically in time.


Alternatively, MQST can be understood from a few-body perspective
\cite{zollner_ultracold_2006,streltsov_role_2007,zollner_excitations_2007,
	zollner_few-boson_2008,zollner_tunneling_2008,
	sakmann_exact_2009,sakmann_quantum_2010,
	cao_interaction-driven_2011,sakmann_universality_2014,	
	murmann_two_2015,liu_two_2015,
	dobrzyniecki_exact_2016,harshman_infinite_2017}
via correlated tunneling
\cite{zollner_excitations_2007,
	zollner_few-boson_2008,zollner_tunneling_2008,sakmann_exact_2009}.
At weak interactions a single-frequency Rabi oscillation 
evolves gradually into a two-mode beating
with characteristic collapse and revival sequences 
\cite{milburn_quantum_1997,raghavan_transitions_1999}.
As interactions become stronger, 
the discrepancy among frequencies increases 
\cite{zollner_few-boson_2008,zollner_tunneling_2008}, 
resulting in a high-frequency mode 
describing first-order tunneling of single atoms 
and a low-frequency mode corresponding 
to a simultaneous co-tunneling of atoms, 
which in fact can be measured in experiments
\cite{winkler_repulsively_2006,folling_direct_2007,
	trotzky_time-resolved_2008,murmann_two_2015}.
Beyond a critical value of interactions 
(correlated with the number of atoms),
the low-frequency mode becomes dominant, realizing MQST, 
which is eventually destroyed for sufficiently long propagation times 
\cite{raghavan_transitions_1999,smerzi_macroscopic_2000,
	zollner_few-boson_2008,zollner_tunneling_2008,
	sakmann_exact_2009,hipolito_breakdown_2010}.


Even though the two-mode model displays good agreement with 
experimental data on BJJ dynamics at short-time scales 
\cite{albiez_direct_2005,levy_c_2007}, 
there is a number of studies reporting discrepancies at longer times,
especially in one-dimensional BJJs featuring enhanced correlations among particles.
For instance, solutions obtained with 
the multi-configuration time-dependent Hartree method for bosons 
(MCTDH \cite{meyer1990multi,beck2000multiconfiguration} 
respectively MCTDHB \cite{alon2008multiconfigurational}),
a variational approach for solving the time-dependent Schr{\"o}dinger equation,
report enhanced inter-band effects \cite{zollner_tunneling_2008,cao_interaction-driven_2011},
universal long-time fragmentation dynamics \cite{sakmann_universality_2014},
conditional tunneling of fragmented pairs \cite{zollner_few-boson_2008} and
MQST being overall reduced by high-order correlations \cite{sakmann_exact_2009}.


An interesting extension of the tunneling problem 
involves mixtures of distinct species,
such as binary Bose  
\cite{kuang_macroscopic_2000,ashhab_external_2002,wen_tunneling_2007,xu_stability_2008,
	mazzarella_atomic_2009,satija_symmetry-breaking_2009,naddeo_quantum_2010,
	pflanzer_material-barrier_2009,pflanzer_interspecies_2010,
	chatterjee_few-boson_2012,chen_impurity-induced_2021}
or Fermi mixtures
\cite{salasnich_macroscopic_2008,salasnich_quantum-tunneling_2010,
	valtolina_josephson_2015,sowinski_diffusion_2016,
	burchianti_connecting_2018,erdmann_correlated_2018,erdmann_phase-separation_2019}
realized by different atoms, isotopes 
or hyperfine states of the same kind of atoms.
In optical double-well traps we can even have spinor condensates 
\cite{ashhab_external_2002,pu_macroscopic_2002,
	mustecaplioglu_tunneling_2005,mustecaplioglu_quantum_2007,
	julia-diaz_josephson_2009,sun_dynamics_2009,wang_spinor_2009},
where spatial tunneling (external Josephson junction) 
is augmented by spin tunneling (internal Josephson junction) \cite{pu_macroscopic_2002}.
The underlying correlations in spin and motional degrees of freedom
realize an atomic analogue of macroscopic quantum tunneling of magnetization (MQTM) 
with potential applications in the framework of magnetic tunneling \cite{chudnovsky_macroscopic_1998}.

The interplay of intra- and inter-species interactions 
greatly impacts the tunneling period of the individual components
and produces novel dynamical regimes, such as
a symmetry-restoring dynamics where the two species avoid each other 
by swapping places between the two wells \cite{ng_quantum-correlated_2003},
a symmetry-broken MQST 
where the two species localize in separate wells 
or coexist in the same well \cite{satija_symmetry-breaking_2009},
and where one component realizes an effective non-rigid \emph{material barrier}, 
see \cite{pflanzer_material-barrier_2009,pflanzer_interspecies_2010,theel_many-body_2021} 
for the definition and use of this concept,
which can interact with tunneling atoms in contrast to a rigid barrier realized by an external trap. 


\colorOne{In the context of binary mixtures, 
a special case of an impurity immersed into a medium
warrants a particular attention.}
A single atom \cite{bausmerth_quantum_2007,fischer_coherent_2008} 
or ion \cite{gerritsma_bosonic_2012,joger_quantum_2014,schurer_impact_2016} 
placed in-between the two wells of a tunneling medium realizes a controlled BJJ.
The internal state of the impurity serves as an additional tunneling channel
and can act as a switch between coherent transport and MQST.
In a similar spirit, the tunneling of an impurity
can be controlled by a background medium \cite{tylutki_coherent_2017,theel_entanglement-assisted_2020}
allowing to change the tunneling period and even to pin it inside the barrier.


In this work we combine several of the above physical insights
to study the transport and tunneling of an impurity in a symmetric double-well
when it becomes immersed into a background of two different bosonic species.
Relevant questions to be addressed are
the possibility to control the state of the impurity via these embeddings,	
the realization of an efficient and at the same time reliable 
transfer of the impurity between the two wells,
as well as the quest for a localization and long-time storage of the impurity.
These questions are not straightforward to answer, 
considering that the build-up of interaction-induced correlations is difficult to predict 
and even more challenging to control
often leading to unexpected outcomes.
Moreover, in order to control the impurity 
we also need to ensure some sort of control over the two majority components.		
Our idea is to initialize the two majority components 
in opposite wells in the MQST regime.
In particular, by manipulating the sign and strength 
of majority-impurity interactions at designated times
we can make each majority species to act either as an attractor or as a repeller for the impurity,
assuming of course that the majority species
stay self-localized for the complete time during the dynamics.

The dynamics is simulated numerically 
by the multi-layer multi-configuration time-dependent Hartree method 
for atomic mixtures (ML-X) \cite{cao2013multi,kronke2013non,cao2017unified},
which takes into account all interaction-induced correlations.	
We work out a successful protocol and analyze the resulting dynamics of, among others, 
the one-body densities to visualize the motion of particles
and quantify the performance of our protocol.	
Furthermore, we investigate the build-up and impact of entanglement for any pair of species,
and employ an effective potential description for the impurity,
which is reminiscent of tunneling in an asymmetric double well
\cite{dounas-frazer_ultracold_2007,carr_dynamical_2010,zollner_tunneling_2008,xu_stability_2008}
where the asymmetry changes over time.


This work is structured as follows. 
In \cref{sec:setup} we introduce our Hamiltonian. 
In particular, we characterize the initial state and
motivate a time-dependent control sequence
of majority-impurity interactions meant to realize a controlled transport.
In \cref{sec:methods} we provide essential details on the numerical approach 
and formulate explicitly our variational ansatz for the many-body state.
The obtained results are described, discussed and analyzed in \cref{sec:results}.
Finally, in \cref{sec:summary} we provide our conclusions and a corresponding outlook.

\section{Setup, Hamiltonian and Propagation Protocol}\label{sec:setup}

\begin{figure}[htb] 
	\centering
	\includegraphics[width=0.45\textwidth]{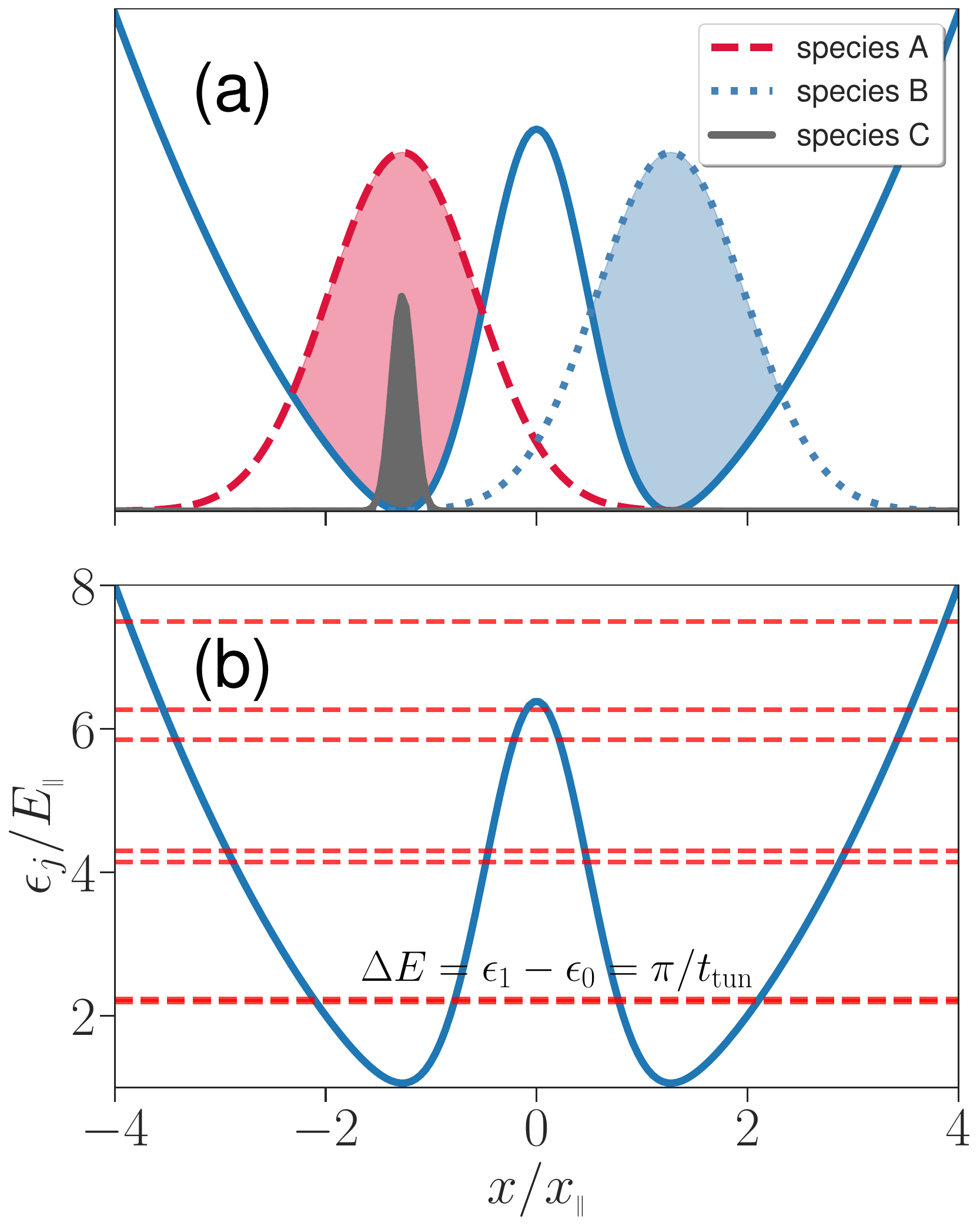}
	\caption{(a) Illustration of the initial ($t=0$) setup for the subsequent dynamics. 
		The colored areas indicate the one-body densities of the different species.
		The majority species $A$ (red) and $B$ (blue) 
		are displayed broader and with a larger maximum as compared to the impurity $C$ (grey),
		indicative of the corresponding particle-number ratios.
		The species $A$ and $C$ start in the left well ($x<0$), 
		whereas the species $B$ is positioned in the right well ($x>0$).
		The interactions among identical bosons are weakly repulsive,
		whereas the inter-species interactions are switched off.
		(b) Spectrum of a single particle (impurity $C$)
		in a symmetric double-well potential, see \cref{eq:Hsigma1}.}
	\label{fig:DWsetup}
\end{figure}

We study a three-component particle-imbalanced mixture.
We assume equal masses $m_{\sigma}=m$ 
with $\sigma \in \{A,B,C\}$ denoting the component label. 
The components $A$ and $B$ have ten bosons each, $N_A=N_B=10$, 
and are referred to as \textit{majority} species,
while the component $C$ is composed of a single particle, $N_C=1$, 
called the \textit{impurity}.
Each species is subject to a one-dimensional double-well confinement
realized as a cigar-shaped harmonic oscillator potential 
($\omega_{\perp} \gg \omega_{\parallel}$)
superimposed with a Gaussian barrier along the longitudinal direction ($x$-axis).
By introducing dimensionless units $E_{\parallel} = \hbar \omega_{\parallel}$ for the energy, 
$x_{\parallel} = \sqrt{\frac{\hbar}{m \omega_{\parallel}}}$ for the length 
and $t_{\parallel} = \frac{1}{\omega_{\parallel}}$ for the time with $\hbar$ being the Planck constant,
the external potential reads
$V_{\rm{dw}}(x)=\frac{1}{2} x^2 + \frac{h}{\sqrt{2\pi}w}e^{-\frac{x^2}{2w^2}}$,
where the width $w=0.5$ and the height $h=8$ of the barrier 
are fixed for the remainder of this work.
The corresponding single-particle energy spectrum 
is depicted in \cref{fig:DWsetup}(b).
Finally, we assume the zero-temperature limit.
Thus, a particle of component $\sigma$ 
interacts with a particle of component $\sigma'$ 
via a s-wave contact-type potential of strength $g_{\sigma \sigma'}$,
which is tunable by Feshbach \cite{chin_feshbach_2010} 
or confinement-induced resonances 
\cite{olshanii_atomic_1998,bergeman_atom-atom_2003,
	haller_confinement-induced_2010}.

Explicitly, the single-species Hamiltonian $\mathcal{H}_{\sigma}$ 
takes the following form:
\begin{flalign}
	\label{eq:Hsigma}
	\mathcal{H}_{\sigma} &= \mathcal{H}^{(1)}_{\sigma} + W_{\sigma}, \\
	\label{eq:Hsigma1}
	\mathcal{H}^{(1)}_{\sigma} &= \sum_{i=1}^{N_{\sigma}} 
	\left(
	- \frac{1}{2} \frac{\partial^2}{(\partial x_i^{\sigma})^2} 
	+ V_{\rm{dw}}(x_i^{\sigma})
	\right), \\
	\label{eq:Hsigma2}
	W_{\sigma} &= g_{\sigma \sigma} 
	\sum_{i<j}^{N_{\sigma}} \delta(x_i^{\sigma}-x_j^{\sigma}),	
\end{flalign}
with $x_i^{\sigma}$ the spatial coordinate 
of the $i$-th particle of component $\sigma$
and $g_{\sigma \sigma}$ the intra-species interaction strength 
among identical particles. 
The triple-mixture Hamiltonian reads:
\begin{equation}
	\label{eq:H}
	\mathcal{H}_t = \sum_{\sigma} \mathcal{H}_{\sigma} + 
	\frac{1}{2} \sum_{\sigma \neq \bar{\sigma}} g_{\sigma \bar{\sigma}}(t)
	\sum_{i}^{N_{\sigma}}\sum_j^{N_{\bar{\sigma}}} 
	\delta(x_i^{\sigma}-x_j^{\bar{\sigma}}),
\end{equation}
with $g_{\sigma \bar{\sigma}}(t)$ the time-dependent interaction strength 
among distinct particles ($\sigma \neq \bar{\sigma}$). 

In the following, we aim to switch between `tunneling' 
and `single-well localized' regimes for the impurity.
To this end, we first initialize our system 
in the ground state of a Hamiltonian $\mathcal{H}_{\rm{rlx}}$ 
(see \cref{ssec:rlx}).
It describes a disentangled ($g_{\sigma \bar{\sigma}} = 0$) mixture,
which is augmented by a species-dependent tilt potential.
In particular, we prepare the two majority species 
at different wells in a self-trapped regime
and trigger tunneling oscillations of the impurity between the two wells,
see \cref{fig:DWsetup}(a).
Subsequently, in \cref{ssec:prop}
we exploit the two spatially-separated majority-species embeddings and
employ a simple time-dependent control sequence 
of the majority-impurity couplings, 
\ie, $g_{AC}(t)$ and $g_{BC}(t)$, 
to either trap the impurity inside one particular well 
or release it to tunnel again. 
The dynamics is then governed by $\mathcal{H}_t$ in \cref{eq:H}.

\subsection{Relaxation}
\label{ssec:rlx}

The initial setup is illustrated in \cref{fig:DWsetup}(a). 
The two majority species are prepared spatially separated on opposite sides 
of the double-well barrier in a self-trapped regime. 
The impurity can be localized in any of the two wells. 
Here, we choose the left well.
Explicitly, we set $g_{\sigma \sigma} = 0.2$ 
and overlay a species-dependent linear tilt 
$V_{\sigma}(x)= -d_{\sigma} \cdot x$ 
with $d_{A}=-d_{B}=d_{C}=0.5$
to energetically favor a particular side of the double-well, \ie,
to account for the `loading' process.		
Thus, species $A$ and $C$ experience a force to the left ($x<0$) 
and species $B$ to the right ($x>0$).
The two majority components 
will act as site-dependent species embeddings for the impurity during propagation,
but for now the inter-species interaction parameters are switched off, 
\ie, $g_{\sigma \bar{\sigma}} = 0$.
The corresponding Hamiltonian reads:
\begin{equation}
	\label{eq:Hrlx}
	\mathcal{H}_{\rm{rlx}} = \sum_{\sigma} \mathcal{H}_{\sigma} + V_{\sigma}(x).
\end{equation}
Our initial many-body state is the ground state of \cref{eq:Hrlx}.
It is obtained using \mlx by time propagating 
the non-interacting ground state of 
$\sum_{\sigma}\mathcal{H}^{(1)}_{\sigma}+V_{\sigma}(x)$ in imaginary time.	

Note that the tilt $|d_{\sigma}|$ 
required to realize a self-trapped regime 
for a single-component condensate 
depends in a non-trivial way on the number of particles 
and the strength of intra-component interactions.
For absent inter-component interactions,
we have verified long-term localization of the two majority species in the initialized wells,
which is a dynamical property we want also to maintain at finite majority-impurity interactions.
That this must be the case is by far not obvious.
What is rather more likely is that
the self-trapping becomes destabilized or even destroyed by the impurity.

Alternatively to the species-dependent tilt potential $V_{\sigma}(x)$
for the initial state preparation 
one can set $g_{AB}$ to be repulsive 
such as to realize a phase-separated state 
between the majority species $A$ and $B$
and then impose a species-independent tilt. 
A sufficiently repulsive $g_{AB}$ can ensure 
that the species $A$ and $B$ stay localized at opposite wells,
whereas the impurity $C$ relocates to a well favored by the chosen tilt.

\subsection{Propagation}
\label{ssec:prop}

Given the initial state of a decoupled mixture ($g_{\sigma \bar{\sigma}}=0$) 
from \cref{ssec:rlx} at $t=0$,
we instantaneously switch off the tilt 
to recreate the symmetric double well, \ie, $d_{\sigma}=0$.
The dynamics is now governed by $\mathcal{H}_{t}$ from \cref{eq:H}.
The majority components become self-trapped 
owing to repulsive intra-species interactions,
whereas the impurity undergoes tunneling.
When the strength of majority-impurity interactions is at zero,
the tunneling period $t_{\rm{tun}}=\pi/\Delta E$ for the impurity
is determined by the energy gap $\Delta E$
between the two lowest eigenstates of \cref{eq:Hsigma1} with $\sigma=C$,
which for the selected double-well in \cref{fig:DWsetup}(b) 
equals $t_{\rm{tun}}=90$ in harmonic units.

\begin{figure}[htb] 
	\centering
	\includegraphics[width=0.45\textwidth]{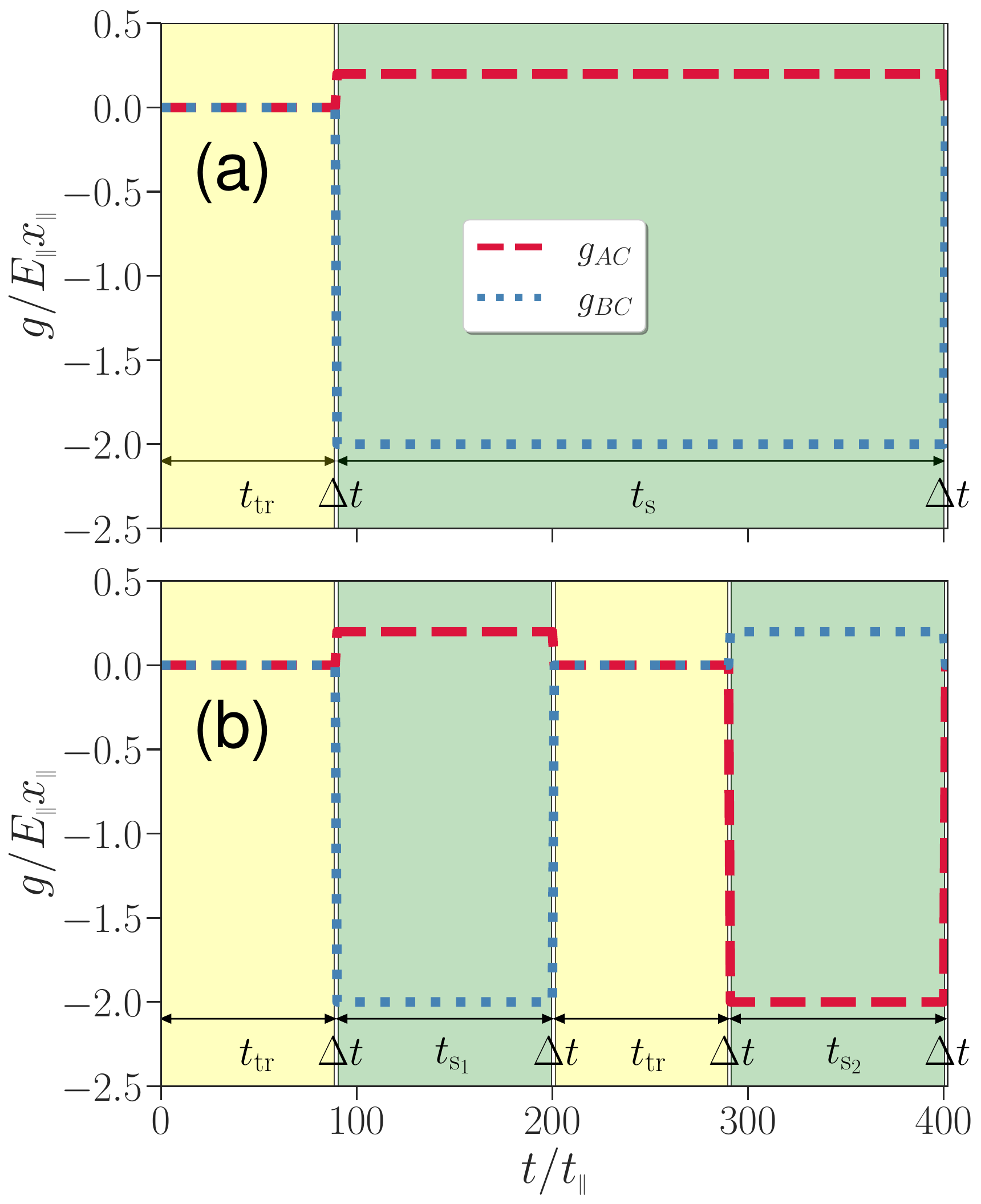}
	\caption{(a) Interaction protocol `transfer-pin-store-unpin'
		characterized by a fixed transfer time $t_{\rm{tr}}=t_{\rm{tun}}-1$, 
		which is determined by the tunneling time $t_{\rm{tun}}/t_{\parallel}=90$ 
		according to the splitting of the lowest doublet in \cref{fig:DWsetup}(b), 
		flexible storage time $t_s$ and short (un)pinning times $\Delta t/t_{\parallel}=1$.
		At $t=0$, the impurity $C$ starts in the left well, occupied by species $A$.
		First, at $0<t<t_{\rm{tr}}$, 
		it tunnels (yellow-shaded area) to the right well, occupied by species $B$.
		Then, at $0<t-t_{\rm{tr}}<\Delta t$,
		it is quickly pinned inside the right well
		by a linear ramp (narrow white-shaded area) of majority-impurity interactions 
		\colorThree{with $g_{AC} \rightarrow g_+$ becoming repulsive and $g_{BC} \rightarrow g_-$ attractive.}
		Note that in the figure the linear ramp resembles a quench due to very short times.
		Once pinned, at $0<t-t_{\rm{tr}}-\Delta t<t_s$, 
		it is stored (green-shaded area) inside the right well, here $t_s/t_{\parallel}=310$.
		Finally, at $0<t-t_{\rm{tr}}-\Delta t-t_s<\Delta t$ it is unpinned by a linear ramp 
		of interactions back to zero.
		(b) Interaction scheme `back-and-forth-transfer' 
		with variable storage times $t_{s_1}$ and $t_{s_2}$.
		At $t=0$, the impurity starts in the left well.
		First, we perform `transfer-pin-store-unpin' to the right, 
		similar to \cref{fig:protocol}(a) 
		except for a different storage time $t_{s_1}/t_{\parallel}=110$.
		Afterwards, we perform `transfer-pin-store-unpin' to the left
		with a storage time $t_{s_2}/t_{\parallel}=110$.
		Note that for the back transfer 
		the interactions have been inverted with 
		\colorThree{$g_{AC} \rightarrow g_-$ becoming attractive and $g_{BC} \rightarrow g_+$ repulsive.}}
	\label{fig:protocol}
\end{figure}

\colorOne{Now, we keep $g_{AB}(t)=0$ 
and control only the majority-impurity interactions $g_{AC}(t)$, $g_{BC}(t)$
such as to transfer the impurity to the opposite side of the double-well 
and freeze it there.
To this end, we devise a simple time-dependent interaction scheme
depicted in \cref{fig:protocol}(a).
It is a four-step procedure which we call the `transfer-pin-store-unpin' protocol,
which is characterized by the following durations: 
the fixed transfer time $t_{\rm{tr}}=t_{\rm{tun}}-1$,
short (un)pin time $\Delta t=1$ and
flexible storage time $t_s$.}

\colorOne{In the first step, the transfer,
the majority-impurity interaction parameters are kept at zero,
which lasts for $t_{\rm{tr}}$ (yellow-shaded area).
During this time the impurity is allowed to freely tunnel.
Once the tunneling to the opposite well is almost accomplished, 
the impurity features a large overlap with the component $B$.
In the second step, the pin, 
we apply a linear ramp within a very short time window $\Delta t$ 
with $g_{AC} \rightarrow g_+$ becoming repulsive and $g_{BC} \rightarrow g_-$ attractive.
The final values of interactions are $g_{+}=0.2$ and $g_{-}=-2$.
Note that in \cref{fig:protocol}(a),
this step resembles a quench (very narrow white-shaded region).
As a result, the impurity is captured by the component $B$
and is prevented from tunneling back. 
In the third step, the storage,
we keep interactions constant for a flexible duration $t_s$ (green-shaded area).
Finally, in the last step, the unpin,
we very quickly ramp down the majority-impurity interactions linearly back to zero,
within a very short time window $\Delta t$  (very narrow white-shaded area).
From there, the impurity resumes its interrupted tunneling.}

%
%
%

\colorOne{To transfer the impurity forth and back, 
we employ the protocol depicted in \cref{fig:protocol}(b).
Essentially, it applies `transfer-pin-store-unpin' from \cref{fig:protocol}(a) two times.
First, we transfer the impurity to the right well 
and hold it there for $t_{s_1}=110$
with species $A$ being repulsive and $B$ attractive. 
Second, we transfer the impurity back to the original well 
and hold it there for $t_{s_2}=110$.
As opposed to the first sequence,
the species $A$ is now attractive and $B$ repulsive.
Note that in the second sequence 
we use the same transfer period $t_{\rm{tr}}$,
even though the state of the impurity is different from the one at $t=0$.}

%
%

Let us note that, in principle, many other protocols
could be imagined and applied.
Indeed, we have explored several other strategies which, however,
turned out to be much less successful.
Nevertheless, we want to give a brief sketch of some alternative protocols
and why they don't work.
First, we investigated much slower linear ramps 
by starting to change the majority-impurity interactions 
right at the start of transfer periods.
As it turns out, already for a forward transfer 
this results in a fraction of the impurity density 
to be left behind, \ie, a sub-optimal transfer,
whereas the majority component, interacting attractively with the impurity,
sustains a sizable decay of self-trapping, 
thereby making it an unreliable container for the impurity during storage times.
Second, we tried to quench the majority-impurity interactions 
at times immediately before and after the storage period,
\ie, infinitely steep ramps.
While the forward transfer was very promising,
the subsequent back-transfer was \colorThree{less efficient 
as compared to the protocol from \cref{fig:protocol}(b)}
in every aspect and, on top of that, extremely sensitive to the particular time 
of the impurity release.

\section{Method and Computational Approach}\label{sec:methods}

To obtain the initial state and to simulate the subsequent dynamics
we need to solve the many-body Schrödinger equations for imaginary time
$\partial_{\tau} \ket{\Psi(t)} = H_{\rm{rlx}} \ket{\Psi(t)}$ and 
for real time $i \partial_{t} \ket{\Psi(t)} = H_{t} \ket{\Psi(t)}$, respectively.
One method is particularly well-tailored to this problem, 
especially in the context of multi-component systems 
of indistinguishable particles,
the multi-layer multi-configuration time-dependent Hartree method 
for atomic mixtures \cite{cao2013multi,kronke2013non,cao2017unified},
usually abbreviated as ML-MCDTHX but here, 
for short, we call it ML-X.

This ab-initio approach expands the many-body wave function in
a finite orthonormal basis whose vectors are \emph{time-dependent} and 
have a product form, properly symmetrized 
to account for the corresponding exchange symmetry of the identical particles.
Both the basis and expansion coefficients are variationally optimized 
to span the relevant part of the full Hilbert space
at each time step of the state evolution.
This allows to reduce the total number of configurations 
as compared to a time-independent basis,
which provides a great boost in convergence 
and makes larger system sizes numerically accessible.
The multi-configuration ansatz takes 
interaction-induced inter-particle correlations into account,
whereas the multi-layer structure 
introduces a hierarchy of Hilbert-space truncations 
by clustering together strongly correlated degrees of freedom (see below).

Our ansatz for a triple mixture has three layers of expansion.
First, we formally group the spatial degrees of freedom 
of indistinguishable particles $\bigcup_i x^{\sigma}_{i}$ 
into three collective coordinates $q^{\sigma}$.
Each $q^{\sigma}$ is then provided 
with a set of $S_{\sigma}$ time-dependent orthonormal 
species wave functions $\Psi_{i}^{\sigma}(q^{\sigma},t)$.
In the first step, we expand our wave function $\ket{\Psi(t)}$ 
according to the following product form:
\begin{equation}
	\label{eq:toplayer}
	\ket{\Psi(t)} = \sum_{i=1}^{S_A} \sum_{j=1}^{S_B} \sum_{k=1}^{S_C}
	A_{ijk}(t) \ket{\Psi_i^{A}(t)} \otimes \ket{\Psi_j^{B}(t)} \otimes \ket{\Psi_k^{C}(t)},
\end{equation}
where $A_{ijk}(t) \in \mathbb{C}$ are time-dependent expansion coefficients.
This partitioning turns out to be particularly useful when correlations among species
are considerably weaker compared to correlations among identical particles.

Next, since a species wave function characterizes identical bosons, 
each of them is expanded in terms of 
symmetrized and normalized product states $\ket{\vec{n}^\sigma(t)}$ 
(so-called permanents encoding that 
$n_i^{\sigma}$ particles occupy 
a time-dependent single-particle orbital $\varphi^{\sigma}_i(x,t)$):
\begin{equation}
	\label{eq:specieslayer}
	\ket{\Psi_i^\sigma(t)} = 
	\sum_{\vec{n}^\sigma|N_\sigma} C_{i,\vec{n}^\sigma}(t) \ket{\vec{n}^\sigma(t)},
\end{equation}
where $C_{i,\vec{n}^\sigma}(t) \in \mathbb{C}$ are time-dependent expansion coefficients, 
$\vec{n}^\sigma|N_\sigma$ restricts the Fock space to configurations 
with a fixed number of particles $\sum_i n_i^\sigma = N_\sigma$,
further truncated to $s_{\sigma}$ single-particle orbitals which can be occupied.
The Fock space dimension is thus given by a binomial coefficient 
$\binom{N_{\sigma}+s_{\sigma}-1}{N_{\sigma}}$.

Finally, applying any of the, in case of analyticity, equivalent time-dependent variational principles
\cite{broeckhove1988equivalence}
leads to a set of coupled time-differential equations 
for $A_{ijk}(t)$, $C_{i,\vec{n}^\sigma}(t)$ and $\varphi^{\sigma}_i(x,t)$.
The single-particle functions are represented 
in a \emph{time-independent} basis
of $s_g$ spatially-localized functions 
$\chi_{\alpha}(x_{\beta})=\delta_{\alpha,\beta}$ (grid DVR) \cite{light2007discrete}:
\begin{equation}
	\label{eq:gridlayer}
	\ket{\varphi_i^{\sigma}(t)} = \sum_{\alpha} d_{i;\alpha}^{\sigma}(t) \ket{\chi_{\alpha}},
\end{equation}
where $d_{i;\alpha}^{\sigma}(t) \in \mathbb{C}$ are time-dependent expansion coefficients.
Note that our grid does not depend on the species label $\sigma$.

\colorOne{The parameter $s_g$ defines the number of grid points to resolve
spatial variations of time-evolving single particle functions.
To fulfill this requirement, we choose an equally spaced grid with $s_g=300$ 
spanning an interval $\left[x_{\rm{min}}, x_{\rm{max}}\right]=\left[-7,7\right]$. 
The parameter $S_{\sigma}$ truncates correlations between \emph{distinct} particles, 
the so-called \emph{inter-species} correlations or entanglement.
For selected physical parameters we find $S_A=S_B=S_C-1=3$ to be suitable 
to faithfully capture the dynamical build-up of entanglement.
The parameter $s_\sigma$ truncates correlations among identical particles,
the so-called \emph{intra-species} correlations or fragmentation.
We find $s_{\sigma}=4$ to be sufficient to account for majority depletion,
which is primarily caused by majority-component interactions of strength $g_{\sigma \sigma}$.
Note that $S_{\sigma}\leq\binom{N_{\sigma}+s_\sigma-1}{N_{\sigma}}$.
Simulations performed with the above choice
of numerical parameters $S_\sigma$, $s_{\sigma}$ and $s_g$
will be referred to as ML-X simulations.
Additionally, let us mention two kinds of approximate solutions.
Setting $S_\sigma=1$ neglects entanglement among species 
and is called a species-mean-field (SMF) ansatz.
Setting $s_\sigma=1$, implying also that $S_\sigma=1$, 
neglects all types of correlations and is known as a mean-field ansatz 
or coupled Gross-Pitaevskii equations (cGPE).}

\section{Results, Analysis and Discussion}\label{sec:results}

\subsection{A single transfer}\label{subsec:1t}

\begin{figure*}[htb] 
\centering
\includegraphics[width=\textwidth]{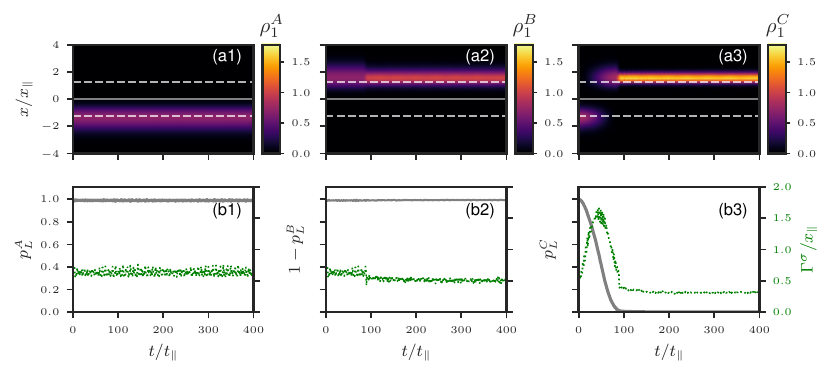}
\caption{Time-evolution of several observables 
	after quenching the tilt of the external potential to zero and, subsequently, 
	following the majority-impurity interaction scheme as depicted in \cref{fig:protocol}(a).
	(a1)-(a3): The one-particle density distribution
	$\rho_1^{\sigma}(x,t)$ of species $\sigma$: 
	$A(B)$ denotes a self-trapped majority component
	while $C$ stands for the impurity.
	The gray solid line at $x=0$ indicates the position of the double-well barrier and
	the gray dashed lines at $x/x_{\parallel}\approx \pm1.27$ the position of the double-well minima.		
	(b1)-(b3): the integrated probability $p_{L}^{\sigma}(t)$ or 1-$p_{L}^{\sigma}(t)$ (gray solid line)
	to find a particle of component $\sigma$ 
	on the left or right side of the double-well potential, respectively,			
	and the standard deviation of the density distribution 
	$\Gamma^{\sigma}(t)$ (green dotted line).} 
\label{fig:rho_hin}
\end{figure*}

\colorThree{First, we apply the interaction protocol from \cref{fig:protocol}(a).}
The goal of this scheme is 
to realize a smooth transfer of an initially localized impurity to the opposite well 
and, subsequently, to store it inside that well for a specified time period $t_s$
while maintaining the shape of the underlying density distribution 
to resemble a Gaussian of a similar width as the initial wave-packet.
The majority components 
are required to remain self-trapped and well-localized during the entire protocol.
It goes without saying that such a transfer 
where the impurity is embedded into separate background majority species
on the left and on the right well
is not only a physically very different situation from the transfer of an isolated single atom
but is also much more difficult to achieve.

In \cref{fig:rho_hin} (a1)-(a3)
we show the time-evolution of one-body densities $\rho_1^{\sigma}(x,t)$ for each species.
In \cref{fig:rho_hin} (b1)-(b3) we present
the corresponding integrated quantities: 
i) $p^{\sigma}_{L}(t)=\int_{-\infty}^0 dx \; \rho_1^{\sigma}(x,t) = 1-p^{\sigma}_{R}(t)$,
which indicates the probability for a particle of species $\sigma$
to be located on the left side \wrt the double-well barrier,
and ii) $\Gamma^{\sigma}(t) = \sqrt{\int dx \; x^2 \rho_1^{\sigma}(x,t) - [\int dx \; x \rho_1^{\sigma}(x,t)]^2}$ 
which is the standard deviation of the corresponding density distribution.

The majority component $A$, see \cref{fig:rho_hin} (a1) and (b1), 
initialized in the left well,
the same as the impurity,
is barely affected by the protocol.	
During the transfer period $t_{\rm{tr}}$, when the majority-impurity interactions are at zero,
the observed dynamics is a result of the initialization procedure, 
namely quenching the tilt of the external potential to zero
triggers low-amplitude high-frequency dipole-like oscillations in the initial density distribution.
\colorOne{Note that this dynamics does 
	not destroy the self-trapping regime of the majority component for long times.}
By the time the majority-impurity interactions are switched on, 
the impurity has tunneled from the left to the right well and
the component $A$, interacting now repulsively with the impurity, 
has no sizable overlap with it during the storage time $t_{s}$ to be noticeably affected.			
Thus, the species $A$ remains self-trapped and localized in the left well the whole time as desired.	
In particular, it acts as a `material barrier' for the impurity
making it energetically unfavorable for the impurity to tunnel back to its initial left well.

The majority component $B$, see \cref{fig:rho_hin} (a2) and (b2), 
initialized in the right well
exhibits a mirror dynamics compared to component $A$ during the transfer time $t_{\rm{tr}}$.
During the storage period $t_s$, when the interactions between 
the component $B$ and the impurity become attractive 
and there is a large overlap between them,
the component $B$ becomes slightly compressed while density fluctuations get reduced.
Importantly, the species $B$ also remains self-trapped and well-localized.	
On top of that, it acts as a `container' for the impurity
preventing it from dispersing within the storage well.

The impurity $C$, see \cref{fig:rho_hin} (a3) and (b3),
first undergoes a free tunneling process during the transfer time $t_{\rm{tr}}$:
the density distribution delocalizes and finally localizes again at the opposite well.
Then, the impurity becomes quickly pinned 
accompanied by an additional compression of the density.
During the storage time $t_s$
it remains highly localized 
and features only minor fluctuations of the mean position and width
reminiscent of sloshing oscillations.

\subsection{Back-and-forth transfer}\label{subsec:2t}

\begin{figure*}[htb] 
	\centering
	\includegraphics[width=\textwidth]{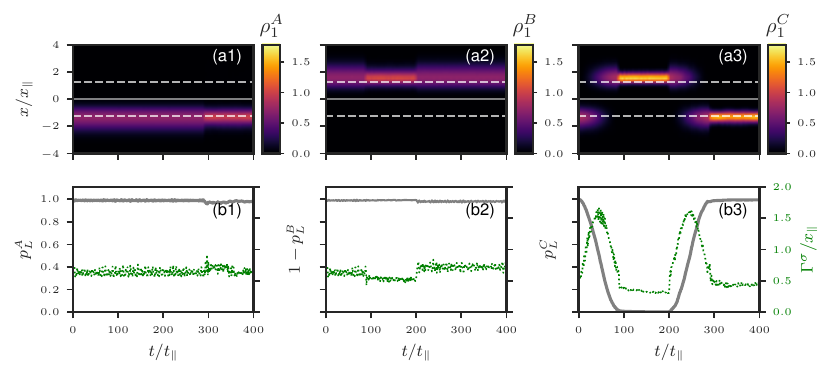}
	\caption{Time-evolution of several observables 
		after quenching the tilt of the external potential to zero and, subsequently, 
		following the majority-impurity interaction scheme as depicted in \cref{fig:protocol}(b).
		(a1)-(a3): The one-particle density distribution
		$\rho_1^{\sigma}(x,t)$ of species $\sigma$: 
		$A(B)$ denotes a self-trapped majority component
		while $C$ stands for the impurity.
		The gray solid line at $x=0$ indicates the position of the double-well barrier and
		the gray dashed lines at $x/x_{\parallel}\approx \pm1.27$ the position of the double-well minima.		
		(b1)-(b3): the integrated probability $p_{L}^{\sigma}(t)$ or 1-$p_{L}^{\sigma}(t)$ (gray solid line)
		to find a particle of component $\sigma$ 
		on the left or right side of the double-well potential, respectively,			
		and the standard deviation of the density distribution 
		$\Gamma^{\sigma}(t)$ (green dotted line).}
	\label{fig:rho_back}
\end{figure*}

\colorThree{Next, we analyze the interaction protocol depicted in \cref{fig:protocol}(b).}
The goal of this scheme is to demonstrate the reverse process, \ie,
that the impurity can be just as smoothly transported back and pinned at the left well,
by applying a `mirror' protocol starting at $t=t_{\rm{tr}}+2\Delta t+t_{s_1}$.
This is by far not obvious, since the many-body wave function has become species-entangled (see later)
as compared to the initial species-disentangled state at zero inter-species interactions.
Importantly, we find (see \cref{subsec:effmodel}) that the storage performance 
of the second sequence (at the left well) is not sizably affected 
by the storage time $t_{s_1}$
of the first sequence (at the right well),
which is yet another benefit of the protocol \colorThree{from \cref{fig:protocol}(b)}
in addition to its simplicity.

The corresponding observables are shown in \cref{fig:rho_back}.
The majority component $A$, see \cref{fig:rho_back} (a1) and (b1), 
is visibly affected by the interaction protocol 
only when it becomes attractive to the impurity, 
namely during the second transfer-and-storage sequence.
At this time interval ($t>291$), 
it sustains very minor density losses to the opposite well, 
which can seen by an overall decrease of $p^A_L$ (gray solid line), 
but remains otherwise well-localized 
as indicated by only minor fluctuations in $\Gamma^A$ (green dotted line).
The majority component $B$, see \cref{fig:rho_back} (a2) and (b2), 
is also visibly affected by the interaction protocol 
only when it is attractive to the impurity, 
namely during the first transfer-and-storage sequence.
After the interactions have been ramped down,
the density sustains slight losses to the opposite well,
but otherwise restores (to a large extent) 
its initial shape from $t=0$.
Overall, both majority components remain
self-trapped and localized as intended.
The impurity $C$, see \cref{fig:rho_back} (a3) and (b3), 
features slightly higher density losses to the opposite well 
during the second storage time
and is on average less compressed.
Nevertheless, the second transfer-and-storage sequence
is just as smooth and stable.

Considering that a back-and-forth transfer
includes a single transfer as its first sequence,
we concentrate on the former from now on.

\subsection{Entanglement measures and analysis}\label{subsec:entanglement}

\begin{figure}[htb] 
	\centering
	\includegraphics[width=0.5\textwidth]{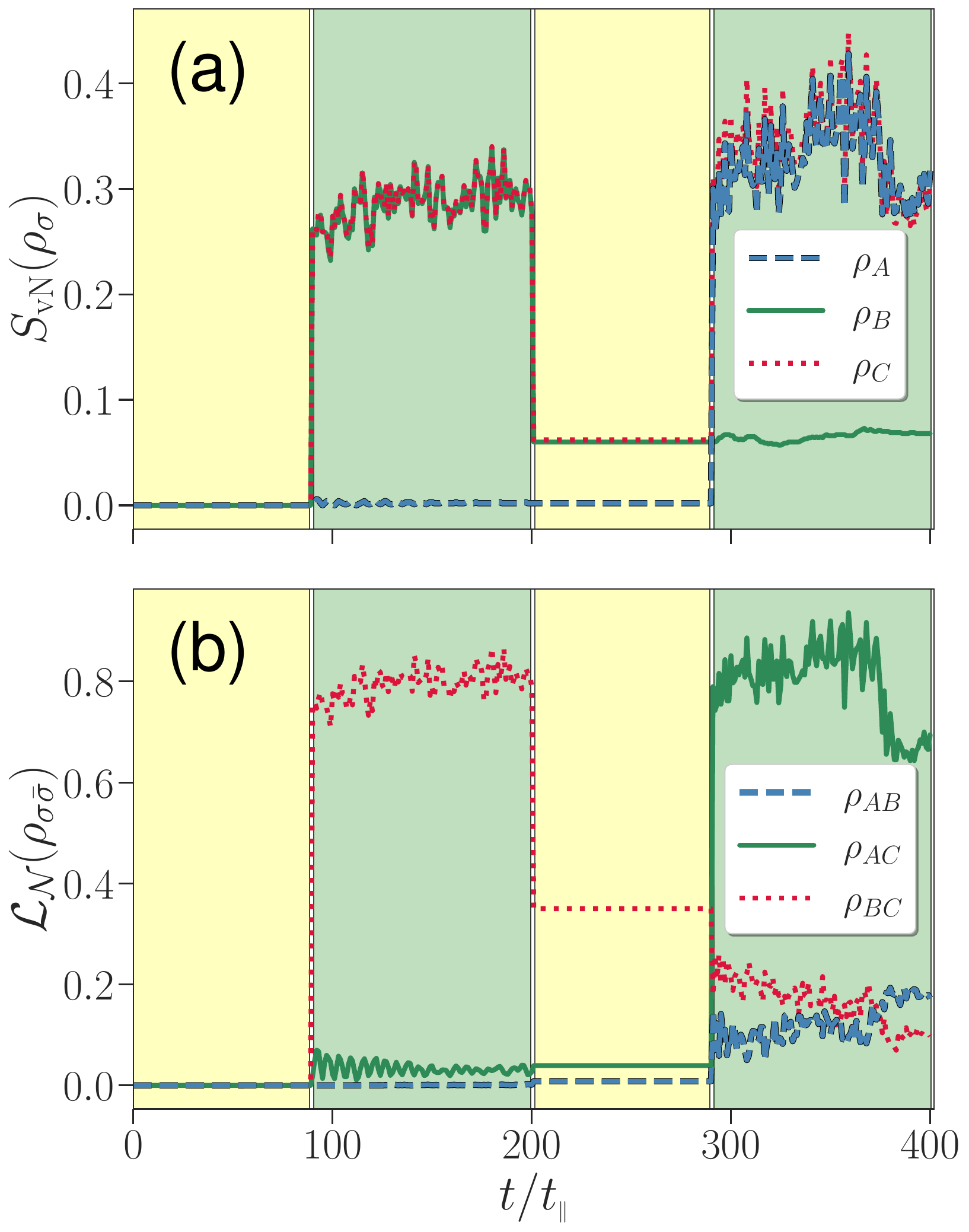}
	\caption{Time-evolution of entanglement measures 
		after quenching the tilt of the external potential to zero 
		and, subsequently, 
		following the majority-impurity interaction scheme 
		as depicted in \cref{fig:protocol}(b).
		(a): the von-Neumann entropy        
		$S_{\rm{vN}}(\rho_{\sigma})$, see \cref{eq:entropy},
		of a reduced single-species density $\rho_{\sigma}$.
		(b): the logarithmic negativity 
		$\mathcal{L}_{\mathcal{N}}(\rho_{\sigma\bar{\sigma}})$, 
		see \cref{eq:logneg},
		for a two-component subsystem described by a mixed state
		$\rho_{\sigma\bar{\sigma}}$ with $\sigma \neq \bar{\sigma}$.}
	\label{fig:negativity}
\end{figure}   

We proceed by analyzing the build-up of entanglement 
for the back-and-forth transfer 
based on the protocol provided in \cref{fig:protocol}(b).
To this end, we employ two measures:
the von-Neumann entropy $S_{\rm{vN}}$ 
and the logarithmic negativity $\mathcal{L}_{\mathcal{N}}$.

The von-Neumann entropy $S_{\rm{vN}}$ characterizes entanglement 
of a bipartite system. 
Our system, however, is tripartite.
To render it bipartite, we partition it into a single-component
and a double-component subsystems.
This gives us three measures 
defined as follows:
\begin{flalign}
	&S_{\rm{vN}}(\rho_{\sigma}) 
	= - \sum_j \lambda_j^\sigma \log(\lambda_j^\sigma),
	\label{eq:entropy} \\
	&\rho_{\sigma} =
	\Tr_{\sigma'\neq\sigma}[\rho] =
	\sum_j \lambda_j^\sigma \ket{\Phi^{\sigma}_j}\bra{\Phi^{\sigma}_j},
	\label{eq:rhosigma}
\end{flalign}
where $\rho_{\sigma}$ 
is the reduced density matrix of a component $\sigma$
obtained from a pure many-body state $\rho$ by tracing out
all particles from other components $\sigma'\neq\sigma$.
Here, it is represented in terms of
natural orbitals $\ket{\Phi^{\sigma}_j}$ (eigenvectors) and
natural populations $\lambda_j^\sigma$ (eigenvalues) 
satisfying $\sum_j \lambda_j^\sigma=1$ and $\lambda_1>\dots>\lambda_{S_{\sigma}}$.
In the absence of entanglement
$\lambda_1=1$ and $S_{\rm{vN}}=0$ vanishes.
For a maximally entangled state all natural populations are the same, \ie, 
$\lambda_j=1/S_{\sigma} \; \forall j$ (see \cref{sec:methods}),
which gives  $S_{\rm{vN}}=\log{S_{\sigma}}$.

\begin{figure*}[htb] 
	\centering
	\includegraphics[width=\textwidth]{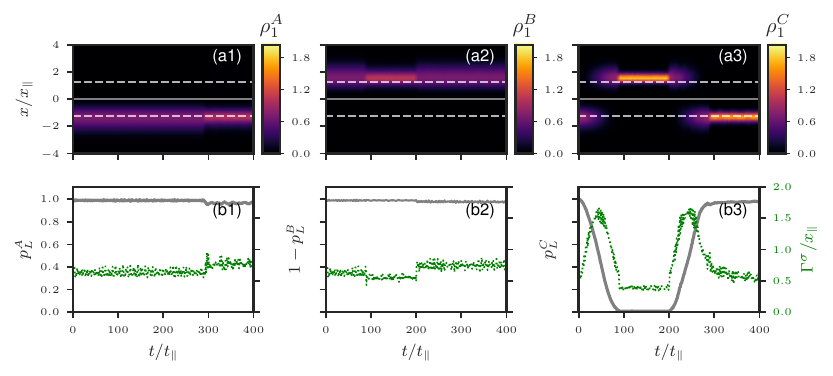}
	\caption{
		Time-evolution of several observables 
		after quenching the tilt of the external potential to zero and, subsequently, 
		following the majority-impurity interaction scheme as depicted in \cref{fig:protocol}(b).
		The physical and numerical parameters are the same as \cref{fig:rho_back}
		except here we neglect the entanglement 
		in the ML-X ansatz ($S_{\sigma}=1$) for the many-body wave function, 
		see \cref{sec:methods}.
		(a1-a3): The one-particle density distribution
		$\rho_1^{\sigma}(x,t)$ of species $\sigma$: 
		$A(B)$ denotes a self-trapped majority component
		while $C$ stands for the impurity.
		The gray solid line at $x=0$ indicates the position of the double-well barrier and
		the gray dashed lines at $x/x_{\parallel}\approx \pm1.27$ 
		the position of the double-well minima.		
		(b1-b3): the integrated probability $p_{L}^{\sigma}(t)$ or 1-$p_{L}^{\sigma}(t)$ (gray solid line)
		to find a particle of component $\sigma$ 
		on the left or right side of the double-well potential, respectively,			
		and the standard deviation of the density distribution 
		$\Gamma^{\sigma}(t)$ (green dotted line).}
	\label{fig:rho_back_smf}
\end{figure*}

The von-Neumann entropy is depicted in \cref{fig:negativity}(a).
Keep in mind that it tells us whether a single component 
is entangled with a pair of other two components.	
In particular, it lacks the ability to resolve
entanglement between any specific two components of a tripartite system.
During the first transfer period, which is free of inter-component interactions,
there is no entanglement as expected.
After the ramp-up, at $t=90$,
the components $B$ and $C$ 
have (individually) accumulated a sizable and comparable amount of entanglement,	
whereas the component $A$ is not (notably) entangled.
This is in accordance with our expectations:
during the ramp-up
$B$ and $C$ feature a large overlap with each other and
almost no overlap with $A$.
Thus, it might be reasonable to assume a product state between subsystems $A$ and $B-C$.	
During the subsequent storage time, when interactions are kept fixed, 
we observe fluctuations of the entropy in components $B$ and $C$.	
After the ramp-down, at $t=200$,
the entropy of $B$ and $C$
has dropped considerably and, 
for the next transfer period at zero majority-impurity interactions, 
becomes frozen.
The component $A$ is still not noticeably affected 
for the same reasons as before.

At the start of the second ramp-up at $t=290$,
the relations become alternated:
$C$ has returned to the left well occupied by $A$, 
such that now $A$ and $C$ become (individually) strongly entangled.
During the subsequent storage time, 
the corresponding entropies undergo larger-amplitude fluctuations
as opposed to $B$ and $C$ in the first storage period.
In addition, they do not exactly match each other.
Regarding the component $B$, 
it preserves the value of entropy accumulated
after the first storage-and-transfer sequence
and features only minor fluctuations
during the second storage period.	

Thus, the evolution of the von-Neumann entropy follows a particular pattern.
It remains frozen at non-interacting transfer times.
For overlapping components it builds up  when interactions are ramped up,
and abruptly decays but stays finite when interactions are ramped down.
Moreover, it features large-amplitude fluctuations
during storage times.	
To resolve to which extent one single component
is entangled with another single component, 
we require a different measure.	

The logarithmic negativity $\mathcal{L}_{\mathcal{N}}$
quantifies pair-wise entanglement 
between two distinct components $\sigma$ and $\bar{\sigma}$,
which are described by a mixed state $\rho_{\sigma\bar{\sigma}}$.
The latter is obtained from a pure many-body state $\rho$ 
by tracing out all particles from the third component
$\sigma'\notin \left\{\sigma,\bar{\sigma}\right\}$:
\begin{equation}
	\rho_{\sigma\bar{\sigma}} =
	\Tr_{\sigma'\notin \left\{\sigma,\bar{\sigma}\right\}} [\rho]
	= \sum_{i,j,k,l} b_{ijkl} 
	\ket{\Psi^{\sigma}_i,\Psi^{\bar{\sigma}}_j}        
	\bra{\Psi^{\sigma}_k,\Psi^{\bar{\sigma}}_l},
	\label{eq:rhosigmabar}        
\end{equation}
here represented in terms of species orbitals 
$\ket{\Psi^{\sigma}_i}$
of the ML-X expansion from \cref{eq:toplayer,eq:specieslayer}.    
The logarithmic negativity $\mathcal{L}_{\mathcal{N}}$ 
depends on the partial transpose $\rho_{\sigma\bar{\sigma}}^{T_{\sigma}}$ 
in the following way:
\begin{flalign}
	&\mathcal{L}_{\mathcal{N}}(\rho_{\sigma\bar{\sigma}}) 
	= \log_2(|\rho_{\sigma\bar{\sigma}}^{T_{\sigma}}|_1) =
	\log_2(1+2 \mathcal{N}),
	\label{eq:logneg} \\
	&\rho_{\sigma\bar{\sigma}}^{T_{\sigma}} =
	\sum_{i,j} b_{kjil} 
	\ket{\Psi^{\sigma}_i,\Psi^{\bar{\sigma}}_j}        
	\bra{\Psi^{\sigma}_k,\Psi^{\bar{\sigma}}_l} =
	(\rho_{\sigma\bar{\sigma}}^{T_{\bar{\sigma}}})^T,
	\label{eq:transpose}       
\end{flalign}
with $|\rho|_1 = \Tr \left\{ \sqrt{\rho^\dagger \rho} \right\}$ 
the trace norm and $\mathcal{N} = \sum_i |\mu_i|$ the negativity,
which is the sum of negative eigenvalues $\mu_i<0$
of $\rho_{\sigma\bar{\sigma}}^{T_{\sigma}}$.
When there is no entanglement between components $\sigma$ and $\bar{\sigma}$, 
$\rho_{\sigma\bar{\sigma}}^{T_{\sigma}}$ is positive semi-definite
and $\mathcal{L}_{\mathcal{N}}=0$ vanishes.
Otherwise, it is positive with larger values indicating a stronger entanglement.  

The logarithmic negativity is depicted in \cref{fig:negativity}(b).
For the first transfer-and-storage sequence ($t<200$),
the high values of entropy for the components $B$ and $C$
can be now indeed attributed
to them being pairwise entangled with each other.	
The entanglement between $A$ and $C$ (green solid line) becomes also more apparent,
though it is still an order of magnitude less than between $B$ and $C$ (red dotted line).
There is no entanglement between $A$ and $B$ (blue dashed line) as expected.

For the second transfer-and-storage sequence ($t>200$)
the distribution of entanglement is less obvious.
Remarkably, during the ramp-up 
we observe a gradual build-up of entanglement between $A$ and $B$.
As they are non-interacting and barely overlap,
it must be mediated by the tunneling impurity.
At the same time, the entanglement between $B$ and $C$ decreases by a similar amount,
which is in accordance with a (roughly) constant entropy of $B$ mentioned before.
The logarithmic negativity between $A$ and $C$ behaves similarly
to the evolution of individual entropies of $A$ or $C$ in \cref{fig:negativity}(a) at $t>291$.	
Thus, the logarithmic negativity provides complementary insights
into the build-up of entanglement in a tripartite system.
To some extent, the entropy of a component $\sigma$
is proportional to a sum of logarithmic negativities involving that component.		

Given that the inter-species entanglement 
is quite sizable during the dynamics,
one might ask what impact it has on the ongoing dynamics, 
in particular the impurity motion.
To this end, in \cref{fig:rho_back_smf} 
we show the previously analyzed observables for the back-and-forth transfer 
assuming now a SMF expansion ($S_{\sigma}=1$) for the many-body wave function, 
as introduced in \cref{sec:methods}.
We remind that this ansatz assumes a single product state in \cref{eq:toplayer},
thus ignoring entirely any inter-species correlations.
Apparently, the majority components are not visibly affected when compared to \cref{fig:rho_back},
whereas the impurity seems to be destabilized 
by the absence of inter-species correlations,
featuring larger fluctuations on the density width, see \cref{fig:rho_back_smf}(b3).
Thus, the build-up of entanglement contributes in a non-trivial way
to a robust transfer and storage of the impurity particle.

\colorOne{Furthermore, a mean-field ansatz ($s_{\sigma}=1$) displays 
a similar dynamics to \cref{fig:rho_back_smf} (see Appendix).
Initially, the majority components are almost condensed.
In the course of the dynamics, they experience only a slight fragmentation ($\sim 3\%$),
which explains the strong similarity between uncorrelated and correlated results.
In this spirit, we have done mean-field simulations 
for $N_A=N_B=20$ with $g_A=g_B=0.1$ and 
$N_A=N_B=50$ with $g_A=g_B=0.04$ (see Appendix).
Again, we observed a very similar dynamics to \cref{fig:rho_back_smf}.
Among the differences, we noticed that during storage 
the width of the impurity density and its fluctuations increase with an increasing number of particles,
which has a negative impact on the storage performance.
However, it might be that correlations will stabilize the impurity,
though we cannot verify it here numerically 
given the increased computational complexity.}

\subsection{Effective potential analysis}\label{subsec:effmodel}

While applying a variational approach, such as ML-X,
to solve the time-dependent Schrödinger equation
turns out to be efficient in terms of sparsity 
of the wave function representation, 
it often comes at the cost of reduced interpretability.
Thus, the variationally optimal single-particle orbitals
can be rarely assigned as eigenstates 
of a single particle in some external potential.
However, having such a picture can be often helpful
to understand some dynamical processes.
One such example is a mean-field picture 
where interacting particles experience the averaged spatial distribution 
of all other particles as an effective external potential
and behave accordingly.

Here, we want to provide a similar viewpoint on the dynamics of the impurity.
To this end,
we are going to decompose the corresponding (one-body) density operator 
$\rho_C \equiv \rho_1^{C}$, see \cref{eq:rhosigma},
into projections $p_j\geq 0$ on single-particle basis states $\left\{\ket{\phi_j}\right\}$,
which gives us a distribution of occupation probabilities over these states.
As our projection basis we choose eigenstates 
of a time-dependent effective Hamiltonian:
\begin{gather}
	(\mathcal{H}^{(1)}_{\sigma} + V_{\rm{ind}}^{\sigma}(t))\ket{\phi^{\sigma}_j(t)}
	= \epsilon^{\sigma}_j(t) \ket{\phi^{\sigma}_j(t)},
	\label{eq:Heff} \\
	V_{\rm{ind}}^{\sigma}(x,t) = \sum_{\sigma'\neq\sigma} 
	N_{\sigma'} g_{\sigma \sigma'}(t)\rho_1^{\sigma'}(x,t), 
	\label{eq:Vind} \\
	p_j^{\sigma}(t) = \braket{\phi^{\sigma}_j(t) | \rho_1^{\sigma}(t) | \phi^{\sigma}_j(t)}, 
	\label{eq:projectors}	
\end{gather}	
where $\rho_1^{\sigma} = \Tr_{N_{\sigma}-1} [\rho_\sigma]$ 
is obtained from $\rho_\sigma$, see \cref{eq:rhosigma}, 
by tracing out all $\sigma$ particles except one,
$\epsilon^{\sigma}_j$ the eigenenergy of $\ket{\phi^{\sigma}_j}$
and $g_{\sigma \sigma'}(t)$ evolving according to
the back-and-forth interaction protocol from \cref{fig:protocol}(b).
We note that $\rho_1^{\sigma}(t)$
is obtained from a correlated many-body state $\rho(t)=\ket{\Psi(t)}\bra{\Psi(t)}$ 
as defined in \cref{sec:methods} with $S_A=S_B=S_C-1=3$ and $s_{\sigma}=4$.

\begin{figure}[htb] 
	\centering
	\includegraphics[width=0.5\textwidth]{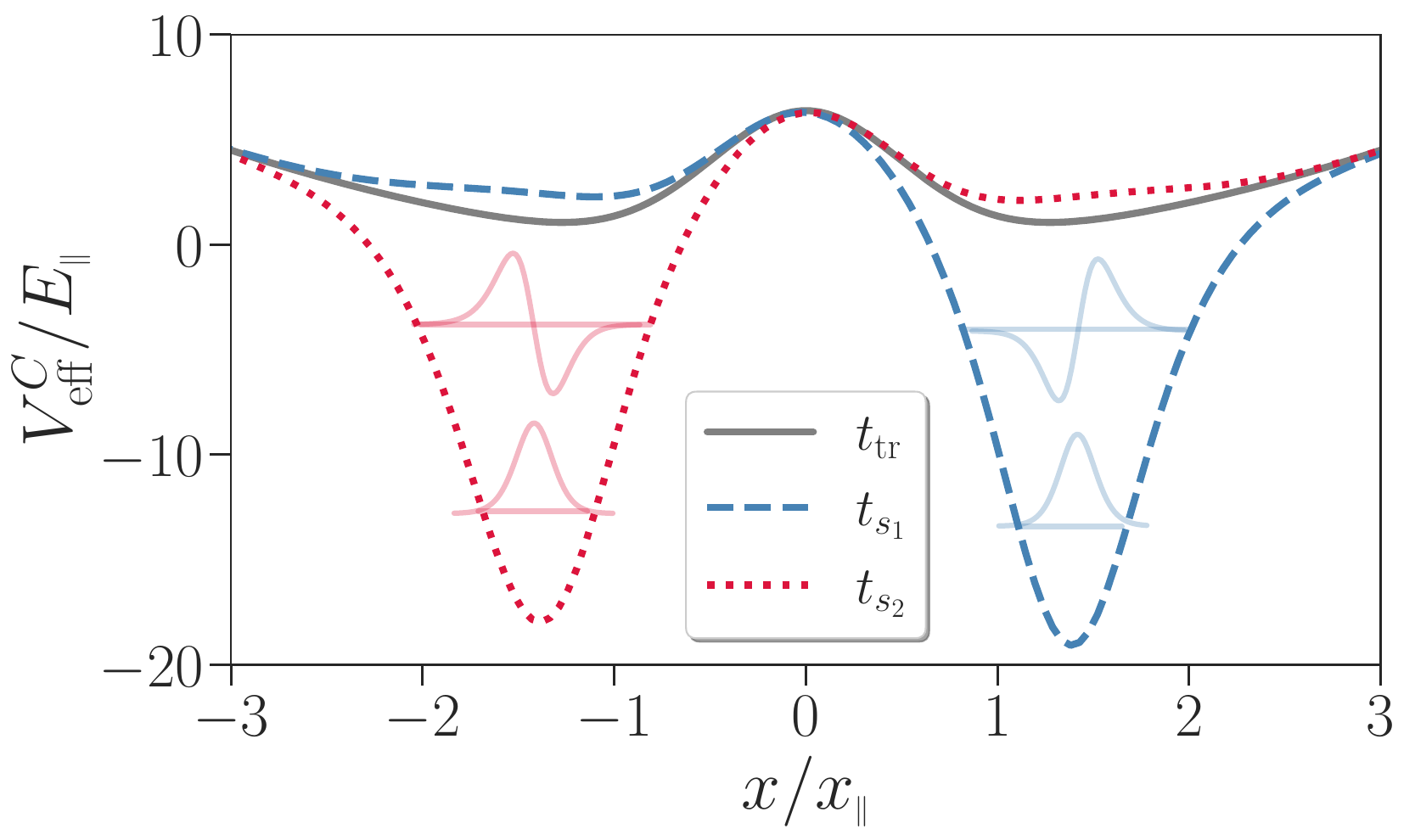}
	\caption{Effective potential $V_{\rm{eff}}^C(t)=V_{\rm{dw}}+V_{\rm{ind}}^C(t)$ 
		for the impurity $C$
		during the back-and-forth interaction protocol 
		from \cref{fig:protocol}(b)
		averaged over selected time intervals:
		transfer period $t_{\rm{tr}}$ (gray solid line),
		first storage period $t_{s_1}$ (blue dashed line) and
		second storage period $t_{s_2}$ (red dotted line).
		The potential is a superposition 
		of an external (static) double-well trap $V_{\rm{dw}}$
		from \cref{fig:DWsetup}(b)
		with a (time-dependent) material barrier and well $V_{\rm{ind}}^C(t)$ from \cref{eq:Vind} 
		which is created by the interchangeably repulsive and attractive majority components.
		The solid red and blue curves indicate 
		the two lowest-energy eigenstates of the corresponding potentials 
		at designated eigenenergies.}
	\label{fig:effpot}
\end{figure} 

The induced potential $V_{\rm{ind}}^{\sigma}$ in \cref{eq:Vind} is
a sum over (time-dependent) one-body densities of the two majority components,
each amplified by the number of particles and 
further modulated by the time-dependent majority-impurity interaction parameter.
We already know from \cref{subsec:2t} 
that in the course of the dynamics the majority components remain self-trapped 
in the initially prepared well. 
Moreover, they take a Gaussian-like shape 
localized at the minimum of the corresponding well
with only small-amplitude fluctuations around that minimum.
Thus, during storage times the repulsive component
represents a potential barrier for the impurity,
thereby decreasing the depth of the corresponding external well,
whereas the attractive component acts as a potential well, \ie, 
it increases the depth of the corresponding external well even further. 
As a result, we get an asymmetric double-well potential, see \cref{fig:effpot}.
Even though this effective potential picture for the impurity is formally related 
to a species-mean-field (non-entangled) ansatz for a triple mixture,
we emphasize that our many-body state $\rho$ 
and the corresponding derived quantities  $\rho_1^{\sigma}$
include inter-species correlations.

\begin{figure}[htb] 
	\centering
	\includegraphics[width=0.5\textwidth]{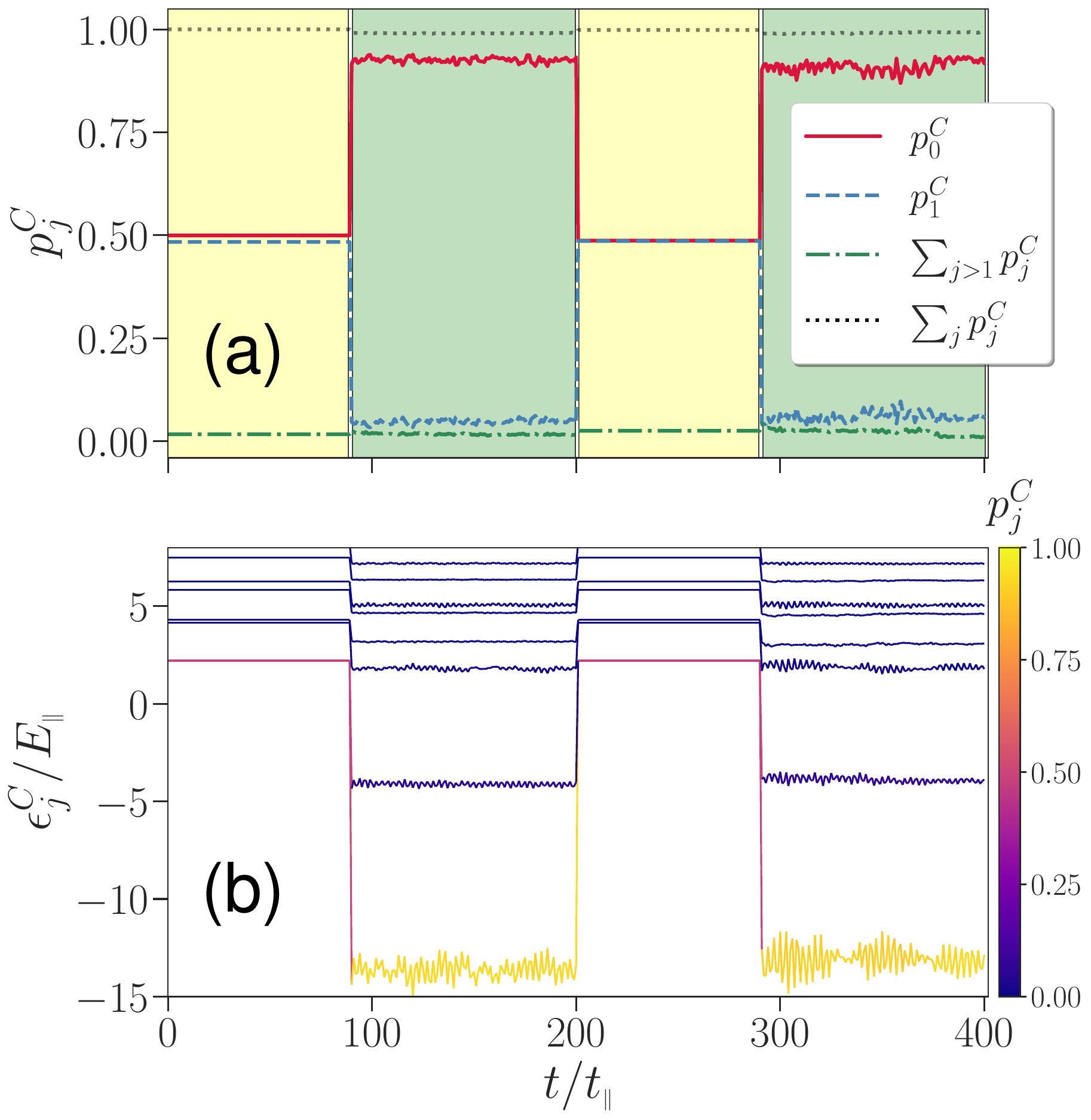}
	\caption{Projectors on instantaneous single-particle eigenstates $p_j^C$ (a)
		and instantaneous eigenenergies $\epsilon_j^C$ (b) 
		of the \emph{time-dependent} effective Hamiltonian from \cref{eq:Heff} for the impurity C,
		which includes time-dependent induced potentials (see \cref{eq:Vind}) 
		created by the majority components $A$ and $B$.
		In (b), each eigenenergy $\epsilon_j^C$ 
		is color-coded with its occupation probability $p_j^C$
		extracted from (a).}
	\label{fig:projectors}
\end{figure}

In \cref{fig:projectors} we show the evolution of probabilities 
$p_j^{C}(t)$ from \cref{eq:projectors} 
for the impurity to occupy the eigenstates 
$\left\{\ket{\phi_j^C(t)}\right\}$ of the Hamiltonian \cref{eq:Heff}
and the corresponding instantaneous eigenenergies $\epsilon^{\sigma}_j(t)$.
Initially, the state of the impurity is 
an almost equal superposition
of the two lowest (quasi-degenerate) eigenstates
of the symmetric double-well potential, see also \cref{fig:DWsetup}(b).
Thus, it tunnels.
After the first ramp-up of interactions at $t=90$,
the right well becomes energetically more favorable, 
and we get an asymmetric double-well, see \cref{fig:effpot} (blue dashed curve).
The impurity still occupies the two lowest eigenstates 
though the weights of occupations are now largely shifted 
in favor of the ground state, which is now a Gaussian localized at the right well.
A slight contribution of the first excited state
explains the high-frequency low-amplitude dipole motion, 
\ie, the left-to-right sloshing, 
of the impurity density inside the right well 
during the storage time.
The low-amplitude fluctuations of occupation probabilities 
are due to interaction with the majority component $B$,
which undergoes a dipole motion inside the right well excited at $t=0$ 
by the instantaneous removal of the external tilt potential.

Once interactions have been ramped down at $t=201$,
we recover the symmetric double-well potential and 
(to a good approximation) the same state composition 
(in terms of amplitudes)
as before the storage sequence.
The impurity resumes
the tunneling motion with the same oscillation frequency.	
The aforementioned sloshing motion of the impurity
impacts the phases of contributing double-well states
and thus also the time it takes to tunnel back.
However, given that the amplitude of sloshing is rather small,
we do not encounter major differences on the transfer time $t_{\rm{tr}}$
upon changing the storage time $t_{s}$.
In other words, 
we can release the impurity at any point in time 
during the storage sequence.
This has been verified numerically for a random sample of storage times
taken in the interval $t_{s_{1}} \in [150,300]$.

Regarding the second transfer-and-storage sequence,
we observe the same patterns except for fluctuations
during the storage time becoming larger.
This might be caused 
by the minor decay of self-trapping in the majority components 
and related density losses to the opposite well, see \cref{fig:rho_back}. 

\section{Conclusions and Outlook}\label{sec:summary}

We have investigated the possibilities 
for a controlled impurity tunneling dynamics in a double-well
containing a mixture of three distinct species.
Building upon insights from the literature,
we prepared two bosonic species of ten atoms each 
in a self-trapped configuration on opposite sides of the double-well barrier
to act as a background for the embedded impurity.
By a suitable manipulation of majority-impurity interactions 
we realized a smooth transport and demonstrated a robust storage of the impurity.		
The study was conducted employing 
the multi-layer multi-configuration time-dependent Hartree method for bosonic mixtures.	

The protocol consists of a sequence of quick ramps of interaction parameters 
and does not require any fine tuning.
To initiate trapping, one majority component is made weakly repulsive and
the other strongly attractive, depending on the storage well.
To initiate transport, interactions are switched off.
The transfer time is determined by the double-well geometry and 
the ramp time needs to be much smaller than the (lowest-band) tunneling time
and the storage time is very flexible (within simulated times).
The protocol is similar in spirit to the pinning procedure in quantum gas microscopy
where one freezes the position of particles by an instantaneous ramp of the lattice depth.

The impurity undergoes a low-frequency large-amplitude dipole oscillation between wells 
during transfer times
and high-frequency small-amplitude dipole motion inside a single well during storage times.
The majority components remain self-trapped 
and perform high-frequency low-amplitude sloshing motion around the double-well minima.
We have analyzed the role of entanglement in terms of the von-Neumann entropy
and the logarithmic negativity.
Our initial state is not entangled.
We find that during ramps the impurity
becomes strongly correlated with the attractive majority component.
Subsequently, the accumulated entanglement undergoes low-amplitude 
modulations during storage times.
After a ramp-down, the entanglement becomes greatly reduced but remains finite.
Interestingly, during the back-and-forth transfer,
we evidenced a build-up of entanglement between the two majority components, 
even though they do not interact and barely overlap.
Apparently, the impurity mediates correlations between spatially-separated majority components.
Furthermore, we compared the correlated many-body dynamics 
to a species-mean-field dynamics, 
which ignores all entanglement effects.
Even though we find a good overall agreement between observables,
there were also sizable discrepancies.
The entanglement has a stabilizing impact on the dynamics
by reducing the amplitude of density fluctuations during transfer and storage times.

Finally, we applied an effective potential picture to describe
the impurity motion as an independent particle evolving
in a time-dependent potential,
which alternates between symmetric and asymmetric double wells.
This potential includes a static double-well
and time-dependent mean-fields produced by the majority particles.
The dynamics is well captured by the two lowest eigenstates of this effective potential.
During transfer times the two eigenstates contribute equally,
and during storage times the ground state dominates with minor fluctuations
caused by oscillations of the mean fields.

Even though the current protocol demonstrates already some very good results,
the underlying minor imperfections might be amplified 
as the number of transfer-and-storage cycles is increased.
The partial decay of self-trapping \colorThree{in the majority species} might be compensated 
by introducing repulsive interactions among majority components
or by changing the strength of intraspecies interactions.
\colorOne{In addition, the entanglement between components 
	was observed to gradually increase with every transfer-and-storage sequence, 
	which might become a limiting factor requiring a disentangling procedure. 
	The latter can be realized by optimizing the ramp times 
	and/or the strength of majority-impurity interactions individually 
	for each transfer-and-storage cycle.}

\colorTwo{Considering that the impurity 
	can be switched between two configurations, left $\ket{L}$ and right $\ket{R}$, 
	the setup might serve as a basic building block of a quantum circuit.
	However, the protocol needs to be modified to
	also include arbitrary superposition states, \ie, $c_L\ket{L} + c_R\ket{R}$.
	To this end, as opposed to the current protocol, 
	we would adapt the transfer times accordingly
	and introduce purely attractive majority-impurity couplings
	to confine each density fraction independently during storage times.
	Finally, to build a quantum circuit, 
	one needs to arrange such qubits in a lattice geometry, 
	\eg, by using arrays of optical tweezers.
	Another interesting topic deserving a thorough investigation
	is the gradual build-up of entanglement 
	between non-interacting majority components,
	mediated through the impurity.}

\begin{acknowledgments}
	This work was supported by the Cluster of Excellence ``Advanced Imaging of Matter" 
	of the Deutsche Forschungsgemeinschaft (DFG) 
	- EXC 2056 -
	Project ID 390715994 and by the
	National Science Foundation under Grant No. NSF PHY-1748958. 		
	J.\ B.\ and M.\ P.\ thank K.\ Keiler for fruitful discussions
	and F.\ Köhler for helping out with ML-X implementations.
	J.\ B.\ and M.\ P.\ thank F.\ Theel 
	for instructing them on the negativity measure.
	
	J.B. and M.P. contributed equally to this work.
\end{acknowledgments}


\appendix

\section{The impact of majority-impurity interaction values.}

\colorThree{Regarding the choice of protocol parameters $g_{\pm}$
for the forward transfer,
we have studied several combinations of parameter values $(g_+,g_-)$
and judged on their performance by evaluating 
the time-averaged probability 
$\tilde{p}_\mathrm{R}^C = \frac{1}{t_s}\int_{0}^{t_s} dt \; p_\mathrm{R}^C (t-t_{\rm{tr}}-\Delta t)$ 
for the impurity to be successfully stored in the right well during the storage time.
The results can be seen in \cref{tab:popul}.
All pairs of considered interaction values perform quite well} 
\colorOne{with only minor differences among them.
	As we were not able to recognize any conclusive trends,
	we have chosen $g_{-}=-2$ and $g_{+}=0.2$ among best performing pairs.}

\begin{table}[h!]
	\centering
	\setcellgapes{3pt}
	\makegapedcells	
	\begin{tabular}{|*{6}{c|}}
		\cline{3-6}
		\multicolumn{2}{c|}{\multirow{2}{*}{$\tilde{p}_\mathrm{R}^C$}} & \multicolumn{4}{c|}{$g_{+}$} \\
		\cline{3-6}
		\multicolumn{2}{c|}{} & $0.1$ & $0.2$ & $0.5$ & $0.7$ \\
		\cline{1-6}
		\multirow{3.5}{*}{$g_{-}$} & $-1.5$ & $0.995 $ & $0.997$ & $ 0.998$ & $0.998$\\
		\cline{2-6}
		& $-2.0$ & $ 0.998$ & $0.998$ & $0.998$ & $0.998$ \\
		\cline{2-6}
		& $-5.0$ & $ 0.997$ & $0.997$ & $0.996$ & $0.996$ \\
		\cline{1-6}
	\end{tabular} 
	\caption{Probability 
		$\tilde{p}_\mathrm{R}^C$
		to find the impurity on the right side of the symmetric double-well potential
		during the storage time $t_s$
		following the interaction scheme \colorThree{from \cref{fig:protocol}(a)} 
		for multiple choices of protocol parameters $g_{+}$ and $g_{-}$.}
	\label{tab:popul}
\end{table}

\colorThree{In a similar way, to select parameters $g_{\pm}$
for the forward and backward transfers,
we evaluated the transfer performance by calculating
the time-averaged probability 
$\tilde{p}_\mathrm{L}^C = \frac{1}{t_{s_2}}\int_{0}^{t_{s_2}} dt \; p_\mathrm{L}^C (t-2t_{\rm{tr}}-3\Delta t - t_{s_1})$ 
for the impurity to be successfully pinned at the left well 
during the second storage sequence.
The results can be seen in \cref{tab:populback}.
As compared to \cref{tab:popul} the storage performance has slightly decreased 
over all parameter pairs, especially at strong attractions $g_-=-5$. 
Our former choice $g_{-}=-2$ and $g_{+}=0.2$ performs comparatively well.}

\begin{table}[h!]
	\centering
	\setcellgapes{3pt}
	\makegapedcells	
	\begin{tabular}{|*{6}{c|}}
		\cline{3-6}
		\multicolumn{2}{c|}{\multirow{2}{*}{$\tilde{p}_\mathrm{L}^C$}} & \multicolumn{4}{c|}{$g_{+}$} \\
		\cline{3-6}
		\multicolumn{2}{c|}{} & $0.1$ & $0.2$ & $0.5$ & $0.7$ \\
		\cline{1-6}
		\multirow{3.5}{*}{$g_{-}$} & $-1.5$ & $0.993$ & $0.994$ & $0.993$ & $0.992$\\
		\cline{2-6}
		& $-2.0$ & $ 0.990$ & $0.992$ & $0.990$ & $0.992$ \\
		\cline{2-6}
		& $-5.0$ & $ 0.987$ & $0.986$ & $0.982$ & $0.983$ \\
		\cline{1-6}
	\end{tabular} 
	\caption{Probability 
		$\tilde{p}_\mathrm{L}^C$
		to find the impurity on the left side 
		of the symmetric double-well potential
		during the storage time $t_{s_2}$
		following the interaction scheme \colorThree{from \cref{fig:protocol}(b)}
		for multiple choices of protocol parameters $g_{+}$ and $g_{-}$.}
	\label{tab:populback}
\end{table}

\section{The impact of particle-number imbalance using a mean-field ansatz.}

\begin{figure*}[htb] 
	\centering
	\includegraphics[width=\textwidth]{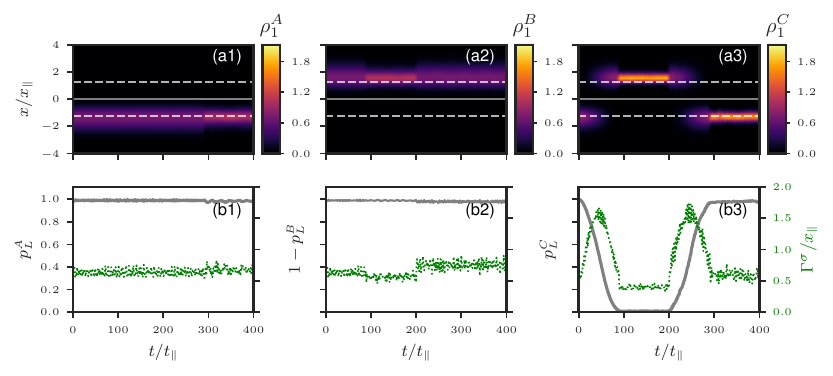}
	\caption{Same as \cref{fig:rho_back_smf} except here we employ a mean-field ansatz. $N_A=N_B=10$ and $g_A=g_B=0.2$.}
	\label{fig:rho_back_mf10}
\end{figure*}

Initially, the majority components are almost condensed.
In the course of the dynamics the degree of fragmentation
gradually increases, 
though the condensed fraction does never drop below $97\%$.
In this spirit, a mean-field ansatz 
might provide a very good qualitative description of the ongoing dynamics.
Indeed, \cref{fig:rho_back_mf10} bears a strong similarity to ML-X simulations from \cref{fig:rho_back}.
There are only slight differences as compared to a species-mean-field ansatz from \cref{fig:rho_back_smf}.

\begin{figure*}[htb] 
	\centering
	\includegraphics[width=\textwidth]{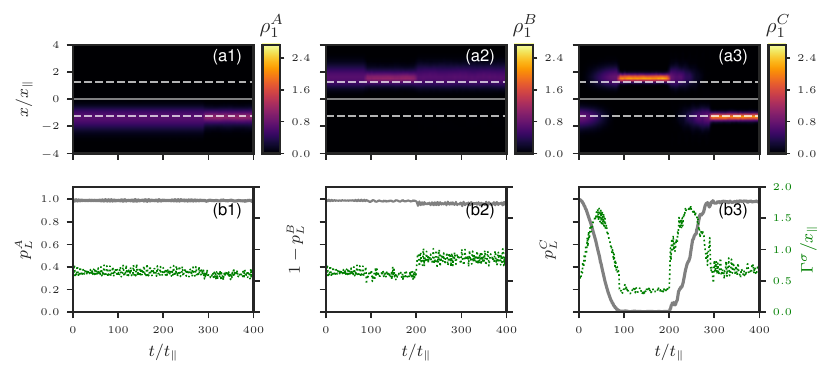}
	\caption{Same as \cref{fig:rho_back_smf} except here we employ a mean-field ansatz. $N_A=N_B=20$ and $g_A=g_B=0.1$.}
	\label{fig:rho_back_mf20}
\end{figure*}

\begin{figure*}[htb] 
	\centering
	\includegraphics[width=\textwidth]{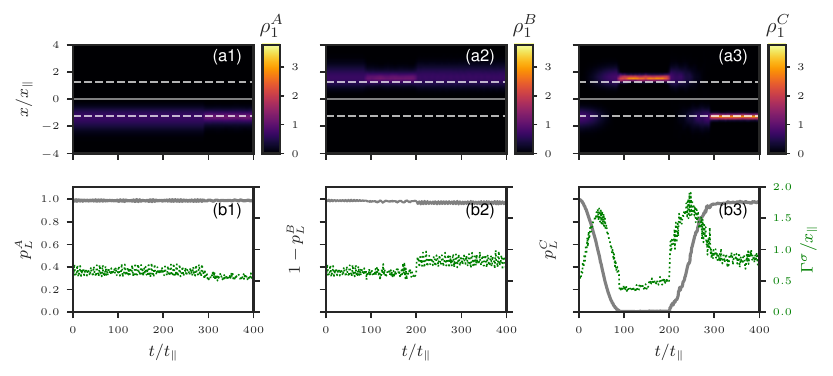}
	\caption{Same as \cref{fig:rho_back_smf} except here we employ a mean-field ansatz. $N_A=N_B=50$ and $g_A=g_B=0.04$.}
	\label{fig:rho_back_mf50}
\end{figure*}

These observations motivate us to employ a mean-field ansatz
to explore the dependence of our protocol on the number of particles in majority components.
In order to maintain the self-trapping regime, we keep $N_{\sigma}g_{\sigma\sigma}$ constant.
In \cref{fig:rho_back_mf20} we show the forth-and-back transfer for $N_A=N_B=20$ and $g_A=g_B=0.1$
and in \cref{fig:rho_back_mf50} for $N_A=N_B=50$ and $g_A=g_B=0.04$.
Qualitatively, the dynamics is similar to \cref{fig:rho_back_mf10}.
Among differences,
the width of the impurity density and its fluctuations increase with increasing number of particles.
This decreases the storage performance.
Nevertheless, the inter-species correlations might stabilize the impurity,
though we cannot verify it with ML-X
owing to the exponential scaling of the Hilbert space dimension with the number of particles.

%
\bibliographystyle{apsrev4-2}

\end{document}